\documentclass[final,5p,times,twocolumn,nopreprintline]{elsarticle}

\usepackage{amssymb}
\usepackage{lipsum}

\usepackage{simpler-wick}
\usepackage{graphicx}  
\usepackage{amsmath}
\usepackage{amsfonts}
\usepackage{amsbsy}
\usepackage{bm}  %
\usepackage{color}
\usepackage{tikz}
\usepackage{xcolor}
\usepackage{hyperref}
\usepackage{float}
\usepackage{xspace}
\usepackage{siunitx}
\usepackage[normalem]{ulem}
\usepackage{balance}
\usepackage{booktabs}
\usepackage{orcidlink}
\usepackage{multirow}

\hypersetup{
    colorlinks,
    linkcolor={red!50!black},
    citecolor={blue!50!black},
    urlcolor={blue!80!black}
}

\usepackage{fancyhdr}
\fancypagestyle{firstpage}{%
	
	\lhead{}
	\rhead{\small INT-PUB-26-042, IPARCOS-UCM-26-058}
}
\newcommand{\bea}{\begin{eqnarray}}
\newcommand{\eea}{\end{eqnarray}}
\newcommand{\beq}{\begin{equation}}
\newcommand{\eeq}{\end{equation}}
\newcommand{\be}{\begin{equation}}
\newcommand{\ee}{\end{equation}}
\newcommand{\bqa}{\begin{eqnarray}}
\newcommand{\eqa}{\end{eqnarray}}

\def\mqo2{{\!\!\!}}

\newcommand*{\Mpi}{\ensuremath{M_\pi}\xspace}

\newcommand{\GeV}{\,\text{GeV}}
\newcommand{\MeV}{\,\text{MeV}}
\renewcommand{\Re}{\text{Re}\,}
\newcommand{\Order}{\mathcal{O}}

\newcommand{\Lb}{\ensuremath{\Lambda_\text{b}}\xspace}
\newcommand{\tLb}{\ensuremath{\tilde\Lambda_\text{b}}\xspace}

\allowdisplaybreaks[4]

\begin{document}

\renewcommand{\theequation}{\arabic{equation}}

\begin{frontmatter}

\title{Naturalness scale inference in chiral perturbation theory and chiral logarithms}

\author[Columbia]{Tara Hassanzadeh\,\orcidlink{0000-0002-3364-158X}}
\author[Bern]{Martin Hoferichter\,\orcidlink{0000-0003-1113-9377}}
\author[UT,ONL]{Lucas Platter\,\orcidlink{0000-0001-6632-8250}}
\author[Madrid]{Jacobo Ruiz de Elvira\,\orcidlink{0000-0001-6089-5617}}
\author[Columbia]{Matthias R.\ Schindler\,\orcidlink{0000-0003-1218-9632}}

\address[Columbia]{Department of Physics and Astronomy, University of South Carolina, Columbia, South Carolina 29208, USA}

\address[Bern]{Albert Einstein Center for Fundamental Physics, Institute for Theoretical Physics, University of Bern, Sidlerstrasse 5, 3012 Bern, Switzerland}

\address[UT]{Department of Physics and Astronomy, University of Tennessee, Knoxville, Tennessee 37996, USA}

\address[ONL]{Physics Division, Oak Ridge National Laboratory, Oak Ridge, Tennessee 37831, USA}

\address[Madrid]{Departamento de F\'isica Te\'orica and IPARCOS, Facultad de Ciencias F\'isicas, Universidad Complutense de Madrid, Plaza de las Ciencias 1, 28040 Madrid, Spain}

\begin{abstract}
We infer naturalness scales of chiral perturbation theory in the mesonic sector using Bayesian techniques that are extended to explicitly take into account chiral logarithms. To demonstrate our approach, we consider the chiral expansion of the pion decay constant $F_\pi$ and the threshold and subthreshold parameters of $\pi\pi$ scattering. We find a strong observable dependence of the inferred naturalness scales, and show that they need not coincide with the pion mass at which the order-by-order hierarchy of the chiral expansion breaks down. 
As an application, we discuss the requirements for lattice-QCD constraints on the two combinations of $\Order(p^6)$ low-energy constants that in the current matching of Roy-equation analyses to chiral perturbation theory need to be estimated from resonance saturation, but could be extracted from the pion-mass dependence of $\pi\pi$ scattering parameters. 
\end{abstract}

\end{frontmatter}

\thispagestyle{firstpage}

\section{Introduction}
\label{sec:intro}

Effective field theories (EFTs) describe low-energy observables as systematic expansions in the ratio of a soft scale, set by the typical momenta and the masses of the light degrees of freedom, to a hard scale $\Lb$, above which the EFT expansion is expected to lose predictive control. While the soft scale is fixed by the process under consideration, the hard scale is usually only known as an order-of-magnitude estimate, typically identified with the mass of the lightest degree of freedom not included explicitly in the EFT. In chiral perturbation theory (ChPT)~\cite{Weinberg:1978kz,Gasser:1983yg,Gasser:1984gg,Leutwyler:1993iq}, the EFT of QCD at low energies, the soft scale is given by the pion momenta and the pion mass, and the standard estimates for $\Lb$ are set by scales such as the $\rho$ meson mass $M_\rho$ and $4\pi F_\pi$, where $F_\pi$ is the pion decay constant, both of order $1\GeV$. A more quantitative characterization of this scale is important for two reasons: first, it gives the denominator of the ratio that controls the uncertainty from the truncation of the chiral expansion at a given order; second, it sets the expected range of pion masses over which ChPT can be used reliably to extrapolate lattice-QCD calculations to the physical point~\cite{FlavourLatticeAveragingGroupFLAG:2021npn,JLQCD:2008zxm,BMW:2013fzj,Boyle:2015exm,CLQCD:2023sdb}, which is still necessary for many calculations of $\pi\pi$ scattering observables~\cite{NPLQCD:2011htk,Dudek:2012gj,ETM:2015bzg,Horz:2019rrn,Mai:2019pqr,Blanton:2021llb,Rodas:2023gma}.

In recent years, Bayesian methods have been developed to quantify truncation uncertainties in EFTs~\cite{Furnstahl:2015rha,Wesolowski:2015fqa,Melendez:2017phj,Wesolowski:2018lzj,Melendez:2019izc}. In this framework, the expansion coefficients are assumed to be of natural size, and the scale $\Lb$ is inferred from the order-by-order pattern of a given observable. Such an analysis has recently been applied to the chiral expansions of the nucleon mass and the nucleon axial coupling~\cite{Ekstrom:2025ost}. Strictly speaking, the scale inferred in this way is a naturalness scale, i.e., the scale for which the expansion coefficients appear to be of natural size. Whether it can be identified with a breakdown scale, i.e., the soft scale at which successive orders become comparable in size, is a separate question, whose answer depends on how the different contributions at a given order combine.

The mesonic sector of ChPT provides a particularly clean setting to address these questions. The pion decay constant and the $\pi\pi$ scattering amplitude are known to two-loop order~\cite{Bijnens:1995yn,Bijnens:1997vq,Bijnens:1998fm,Bijnens:2006zp,Niehus:2020gmf}. The next-to-leading-order (NLO), $\Order(p^4)$, low-energy constants (LECs) $l_i$ are comparatively well constrained~\cite{Ananthanarayan:2000ht,Colangelo:2000jc,Colangelo:2001df,Caprini:2011ky,FlavourLatticeAveragingGroupFLAG:2021npn}, whereas the next-to-next-to-leading-order (NNLO), $\Order(p^6)$, LECs $r_i$ are poorly known and commonly estimated by resonance saturation~\cite{Ecker:1988te,Bijnens:1997vq,Colangelo:2021moe}. In addition, the pion-mass dependence of the observables is accessible in lattice calculations.

However, chiral logarithms appear already at NLO and double logarithms at NNLO. Some of the NNLO logarithmic coefficients are large, with their net effect at the physical point moderated by cancellations whose balance changes with the pion mass~\cite{Colangelo:1995np,Bijnens:1997vq,Girlanda:2001eu}, so that the convergence pattern at the physical point does not by itself guarantee convergence at larger pion masses. 
 In the pointwise analysis applied in the single- and few-nucleon sectors~\cite{Melendez:2017phj,Melendez:2019izc,Ekstrom:2024dqr,Ekstrom:2025ncs,Ekstrom:2025ost}, the dependence on the soft scales is made explicit only through powers of their ratio to $\Lb$, and the coefficients are extracted from the order-by-order corrections at a few selected values of the light scales. The chiral logarithms introduce an additional pion-mass dependence into the coefficients themselves, so that the inferred scale depends on the pion masses at which they are extracted.

In this Letter, we develop an extension of the method that treats the logarithmic and non-logarithmic terms at each order as separate coefficients, which removes the dependence on the choice of pion masses at which the coefficients are evaluated. As we show, this construction amounts to imposing naturalness on the renormalized coefficients at the scale $\mu=\Lb$. We also quantify the sensitivity of the inferred scales to the renormalization-scale prescription and to the treatment of the leading logarithms. After summarizing the ChPT input in Sec.~\ref{sec:chpt}, the formalism for the naturalness scale inference is introduced in Sec.~\ref{sec:inference}, with details provided in \ref{app:threshold-equations} and  \ref{app:bayes}, respectively.

We focus the phenomenological application in Sec.~\ref{sec:results} on a set of representative observables, the pion decay constant $F_\pi$, for which the chiral expansion is particularly well behaved, as well as $\pi\pi$ threshold and subthreshold parameters, which are more strongly affected by chiral logarithms and probe different kinematic regions. The latter proved critical in the pion--nucleon sector~\cite{Hoferichter:2015tha,Hoferichter:2015hva,Siemens:2016jwj}, suggesting that also in the $\pi\pi$ case subthreshold parameters should exhibit better convergence properties given the distance to the unitarity cuts. 
In each case, we test whether the inferred naturalness scale coincides with the pion mass at which the order-by-order hierarchy breaks down and assess the sensitivity to the $r_i$ by comparing two different resonance-saturation inputs~\cite{Bijnens:1997vq,Colangelo:2021moe}.

The most precise phenomenological information on $\pi\pi$ scattering parameters derives from the matching of Roy-equation analyses~\cite{Ananthanarayan:2000ht,Garcia-Martin:2011iqs,Caprini:2011ky} to ChPT. This matching is discussed in great detail in Ref.~\cite{Colangelo:2001df}, where it is set up in a way that minimizes the sensitivity to higher chiral orders and unknown LECs as far as possible.  However, two linear combinations of $r_i$ cannot be determined from the matching, and need to be estimated from resonance saturation, which, as shown in Ref.~\cite{Colangelo:2021moe}, is subject to sizable SU(3) uncertainties that affect especially the LEC $f_\chi$. In this Letter, we consider both variants (Sets~I and~II in Sec.~\ref{ssec:lecs}) to infer the respective naturalness scales in the chiral expansion of the $\pi\pi$ scattering parameters, from which we conclude in Sec.~\ref{sec:lattice} the requirements for lattice-QCD calculations to constrain these LECs from the pion-mass dependence.  We conclude with a summary in Sec.~\ref{sec:summary}.

\section{Mesonic ChPT and \texorpdfstring{$\pi\pi$}{pipi} observables}
\label{sec:chpt}

\subsection{Conventions}
\label{ssec:conventions}

In mesonic ChPT, observables are expanded in powers of a generic soft scale $p$, which stands for both the pion momenta and the pion mass, with the leading order (LO) at $\Order(p^2)$, NLO at $\Order(p^4)$, and NNLO at $\Order(p^6)$. For the static quantities considered in this work, the expansion proceeds in powers of
\beq
\label{eq:xi}
\xi=\frac{\Mpi^2}{(4\pi F)^2},
\eeq
supplemented by the chiral logarithms
\beq
L_\mu=\log\frac{\Mpi^2}{\mu^2},
\eeq
where $\mu$ denotes the renormalization scale. Throughout, we express all quantities in terms of the pion decay constant in the chiral limit, $F$, and the pion mass $\Mpi$, i.e., the full pion mass including its quark-mass corrections, which introduce the dependence on $l_3^r$~\cite{Gasser:1983yg}. The physical value of the pion mass is denoted by $\Mpi^\text{phys}$. The use of $F$ renders the pion-mass dependence at each order explicit, whereas the transition $F\to F_\pi$ reshuffles pion-mass dependence among different orders~\cite{Niehus:2020gmf}. For the pion mass, there is no analogous motivation to retain the LO expression $M^2=2B\hat m$, since $\Mpi$ is the quantity at which the expansion is evaluated, e.g., in lattice-QCD calculations. We note that at the physical point the convergence of the $\pi\pi$ threshold parameters has been observed to improve when the amplitude is expressed in terms of $F_\pi$ instead~\cite{Bijnens:1997vq,Colangelo:2001df}. Since this concerns the organization of higher-order terms rather than a property of the EFT, we do not pursue this possibility here.

We follow the conventions of Refs.~\cite{Gasser:1983yg,Bijnens:1997vq,Bijnens:1999hw,Niehus:2020gmf} for the NLO LECs $l_i^r(\mu)$ and the NNLO LECs $r_i^r(\mu)$. The scale-independent LECs $\bar l_i$ are defined by
\beq
\label{eq:lbar}
l_i^r(\mu)=\frac{\gamma_i}{32\pi^2}\bigg[\bar l_i+\log\frac{(\Mpi^\text{phys})^2}{\mu^2}\bigg],
\eeq
with $\gamma_1=1/3$, $\gamma_2=2/3$, $\gamma_3=-1/2$, and $\gamma_4=2$. When varying the pion mass, the $l_i^r(\mu)$ are kept fixed, so that the pion-mass dependence beyond the explicit powers of $\xi$ resides entirely in the chiral logarithms $L_\mu$. Equivalently, Eq.~\eqref{eq:fpi_corrections} and the expressions in \ref{app:threshold-equations} are written in terms of $\bar l_i(\Mpi)=\bar l_i-\log\big[\Mpi^2/(\Mpi^\text{phys})^2\big]$, i.e., Eq.~\eqref{eq:lbar} with $\Mpi^\text{phys}$ replaced by $\Mpi$, which carry the remaining pion-mass dependence.

\subsection{Pion decay constant}
\label{ssec:fpi_chpt}

The chiral expansion of $F_\pi$ reads~\cite{Gasser:1983yg,Burgi:1996qi,Bijnens:1997vq}
\begin{equation}
F_\pi=F+F_4+F_6+\Order(p^8),
\label{eq:fpi_expansion}
\end{equation}
where $F_4=F \xi \bar l_4(\Mpi)$ and
\begin{align}
\frac{F_6}{F}
&=\xi^2\bigg[r_F^r-\frac{13}{192}-\frac{1}{12}\Big(\bar l_1+4\bar l_2-6\bar l_3\bar l_4\Big)\notag\\
&+\frac{1}{6}\Big(12+7\bar l_1+8\bar l_2+3\bar l_3-3\bar l_4\Big)L_\mu+\frac{5}{4}L_\mu^2\bigg],
\label{eq:fpi_corrections}
\end{align}
with $\bar l_i\equiv\bar l_i(\Mpi)$.
The NNLO LEC $r_F^r$ is not known, and we set $r_F^r(\mu)=0$ at the resonance-saturation scale $\mu=770\MeV$, see Sec.~\ref{ssec:lecs}, with its value at other scales determined by the renormalization-group (RG) running.

\subsection{\texorpdfstring{$\pi\pi$}{pipi} scattering}
\label{ssec:pipi}

In the isospin limit, the $\pi\pi$ scattering amplitude is determined by a single function $A(s,t,u)$~\cite{Weinberg:1966kf,Gasser:1983yg}, in terms of which the $s$-channel isospin amplitudes read
\begin{align}
T^0(s,t,u)&=3A(s,t,u)+A(t,s,u)+A(u,t,s),\notag\\
T^1(s,t,u)&=A(t,s,u)-A(u,t,s),\notag\\
T^2(s,t,u)&=A(t,s,u)+A(u,t,s),
\end{align}
where $s$, $t$, $u$ are the Mandelstam variables, with $s+t+u=4\Mpi^2$. The amplitude is known to $\Order(p^6)$~\cite{Bijnens:1995yn,Bijnens:1997vq}. At this order, it depends on $l_{1\text{--}4}^r$, $r_{1\text{--}6}^r$, and $r_F^r$ (if expressed in terms of $F$).

We consider two classes of observables. The threshold parameters are defined by the expansion of the partial waves $t_\ell^I(s)$ around threshold,
\beq
\label{threshold_expansion}
\Re t_\ell^I(s)=\hat q^{2\ell}\Big[a_\ell^I+b_\ell^I\hat q^2+\Order(\hat q^4)\Big],\qquad
\hat q^2=\frac{s}{4\Mpi^2}-1,
\eeq
where $a_\ell^I$ and $b_\ell^I$ denote the (dimensionless) scattering lengths and effective-range parameters, respectively. We consider the $S$-wave parameters $a_0^0$, $b_0^0$, $a_0^2$, $b_0^2$, and the $P$-wave scattering length $a_1^1$. In addition, we consider the subthreshold parameters $c_0^{0}$ and $c_1^{0}$, defined by the expansion of the isospin amplitudes around the symmetric point $s=t=u=4\Mpi^2/3$, and the position $s_A$ of the Adler zero of the $I=0$ $S$-wave. The subthreshold region lies in the interior of the Mandelstam triangle, farther from the unitarity cuts than the threshold region, so that the chiral expansion is expected to converge better there, as exploited in the matching of Roy--Steiner equations for pion--nucleon scattering to ChPT~\cite{Hoferichter:2015tha,Hoferichter:2015hva,Siemens:2016jwj}. The precise definitions and the explicit chiral expansions are collected in \ref{app:threshold-equations}.

\subsection{Low-energy constants}
\label{ssec:lecs}

For the NLO LECs, we use $\bar l_1=-0.4$ and $\bar l_2=4.3$ from the matching of two-loop ChPT to the Roy-equation solution of Refs.~\cite{Ananthanarayan:2000ht,Colangelo:2001df}, as well as $\bar l_3=3.4$ and $F_\pi/F=1.07$, consistent with Refs.~\cite{FlavourLatticeAveragingGroupFLAG:2021npn,MILC:2010hzw,ETM:2010cqp,Beane:2011zm,Borsanyi:2012zv,BMW:2013fzj,Boyle:2015exm}. With $F_\pi=92.3\MeV$~\cite{ParticleDataGroup:2026mpi}, we then fix $\bar l_4=4.27$ so that Eq.~\eqref{eq:fpi_expansion} at NNLO, with $r_F^r=0$, reproduces $F_\pi$. 
Our naturalness analysis only considers the central values of the LECs without propagating their
errors, since the main uncertainty arises from the NNLO LECs $r_i$, which are much less constrained. We estimate them by resonance saturation~\cite{Ecker:1988te,Bijnens:1997vq}, where the exchange of vector mesons yields
\begin{align}
r_1^r&=-16\alpha-16\beta-4\gamma, &
r_3^r&=-7\alpha-3\beta, &
r_5^r&=\frac{3}{4}\alpha,\notag\\
r_2^r&=20\alpha+16\beta+3\gamma,
& r_4^r&=\alpha+\beta,& 
r_6^r&=\frac{1}{4}\alpha,
\label{eq:ri_res}
\end{align}
at $\mu=770\MeV$, with
\beq
\alpha=\frac{g_V^2F_\pi^2}{M_V^2},\quad
\beta=\frac{4\sqrt{2}g_Vf_\chi F_\pi^2}{M_V^2},\quad
\gamma=\frac{32f_\chi^2F_\pi^2}{M_V^2},
\eeq
and $M_V=775.26\MeV$. The couplings $g_V$ and $f_\chi$ are determined from the decays $\rho\to\pi\pi$ and $K^*\to K\pi$~\cite{Bijnens:1997vq,Colangelo:2021moe}. We consider two sets obtained from the same decays, using either $F_K$ or $F_\pi$ in the $K^*\to K\pi$ width: Set~I, $g_V=0.084$ and $f_\chi=0.0025$, includes SU(3) corrections by means of $F_K$~\cite{Colangelo:2021moe}, whereas Set~II, $g_V=0.091$ and $f_\chi=-0.033$, uses $F_\pi$ and corresponds to the standard values of Ref.~\cite{Bijnens:1997vq}. This choice mainly affects $f_\chi$, which has a sizable impact on $r_{1\text{--}4}$, see Table~\ref{tab:ri}.

\begin{table}[t]
\caption{NNLO LECs $r_i^r(\mu)$ at $\mu=770\MeV$ in units of $10^{-4}$ from resonance saturation, Eq.~\eqref{eq:ri_res}, for the two sets of vector-meson couplings.}\vspace{0.2cm}
\renewcommand{\arraystretch}{1.3}
\label{tab:ri}
\centering
\scalebox{0.9}{
\begin{tabular}{lrrrrrr}
\toprule
 & $r_1$ & $r_2$ & $r_3$ & $r_4$ & $r_5$ & $r_6$\\
\midrule
Set I~\cite{Colangelo:2021moe}  & $-18.8$ & $22.8$ & $-7.5$ & $1.2$ & $0.75$ & $0.25$\\
Set II~\cite{Bijnens:1997vq} & $-0.01$ & $-0.23$ & $-1.0$ & $-1.2$ & $0.88$ & $0.29$\\
\bottomrule
\end{tabular}}
\end{table}

Finally, at fixed pion mass the $\Order(p^6)$ amplitude depends on the $l_i$ and $r_i$ only through a limited number of combinations of its polynomial part, fewer than the number of LECs, so that shifts in the $\bar l_i$ can be compensated by the $r_i$. Within $\pi\pi$ scattering, this $l_i\leftrightarrow r_i$ degeneracy can only be lifted by the pion-mass dependence, see Sec.~\ref{sec:lattice}.

\section{Naturalness-scale inference}
\label{sec:inference}

\subsection{Pointwise analysis}
\label{sec:pointwise}

The EFT expansion for an observable $X$ can be written as
\begin{equation}\label{eq:EFTexpand}
  X=X_\text{ref}\sum_{i=0}^{\infty} c_i Q^i,\qquad Q= \frac{x}{\Lb},\qquad x\in \big\{p,\Mpi\big\},
\end{equation}
where $X_\text{ref}$ is a reference scale, $p$ is the typical momentum, and $\Mpi$ the mass of a light degree of freedom, with both $p$ and $\Mpi$ assumed much smaller than the breakdown scale $\Lb$.
The dimensionless coefficients $c_i$ determine the convergence behavior of the expansion provided $Q$ is known.
Because the light scales can enter $X$ not just through powers of $Q$, the coefficients $c_i$ in Eq.~\eqref{eq:EFTexpand} also depend on $x$.
The pointwise approach infers $\Lb$ from the order-by-order coefficients at selected $x$ values.
The results of the pointwise inference are therefore dependent on the choice of $x$ values at which the analysis is performed.
The pointwise approach also ignores correlations among the $c_i$ at different values of $x$. Gaussian processes can be used to deal with these correlations \cite{Melendez:2019izc}, especially if the exact dependence of the $c_i$ on the light scales is not known.

In mesonic ChPT, observables have an expansion analogous to Eq.~\eqref{eq:EFTexpand}. 
However, only even orders of $Q$ contribute to the expansion. Further, the dependence of the coefficients $c_i$ on the pion mass is known to originate in the chiral logarithms.
For the static quantities considered here, the chiral expansions can therefore be written as
\begin{equation}
    X = X_\text{ref}\sum_{i=0}^{\infty} d_i(\Mpi) q^i,\qquad q=\frac{\Mpi^2}{\Lb^2},
\end{equation}
where each coefficient $d_i$ has an additional expansion of the form
\begin{equation}
\label{eq:dij}
    d_i = \sum_{j=0}^{i} d_{i,j}\, L_\mu^j.
\end{equation}
While the complete $d_i$ at a given order is $\mu$-independent, the $d_{i,j}$ can depend on $\mu$ to cancel the logarithmic dependence.
If the LO expression for $X$ is used as the reference scale $X_\text{ref}$, $d_0=1$ and the logarithmic mass dependence appears in the first nontrivial correction at NLO.
With $\mu$ taken fixed at a value relatively close to the assumed breakdown scale (a typical choice is $\mu=M_\rho$), the chiral logarithms become significant for small pion masses.
In the pointwise analysis, which relies on the coefficients being of natural size, this is reflected in a dependence of the results on the choice of pion masses used in the analysis.

We illustrate this issue for $F_\pi$, Eq.~\eqref{eq:fpi_expansion}, using the chiral-limit value $F$ as the reference scale.
As in previous analyses involving baryons~\cite{Ekstrom:2025ost}, we assume that the coefficients $d_i$ are of natural size and assign a common normal prior, independent and identically distributed (iid),
\begin{equation}
    d_i \overset{\text{iid}}{\sim} \mathcal{N}(0,\bar{c}^2),
\end{equation}
where the variance $\bar{c}^2$ is assigned a conjugate inverse-chi-squared hyperprior,
\begin{equation}\label{eq:hyperprior}
    \bar{c}^2\sim \chi^{-2}(\nu_0,\tau_0^2).
\end{equation}
With the choice $\nu_0=2, \tau_0^2=1$, 67\% of the probability for $\bar{c}^2$ falls into the range $[1/3,3]$. We also assign a log-uniform prior 
to $\Lb$ over the interval $[0.01, 2]$\GeV; changing the interval to  $[0.01, 16]$\GeV~does not impact the results at the reported level of precision.

Table~\ref{tab:PointVsLog} shows representative results of pointwise analyses of the posterior probability distribution for $\Lb$ at NNLO using two pion masses $\{M_\pi^\text{low},M_\pi^\text{high}\}$ (we have performed the analyses for additional pion-mass pairs as well).
The results, even at NNLO, are sensitive to the choice of the masses, and this dependence is particularly strong for $M_\pi^\text{low}$, reflecting the mass dependence of the coefficients through the chiral logarithms.

\subsection{Log-aware analysis}
\label{sec:logaware}

We continue with the pion decay constant as our example to introduce an improved approach, to which we refer  as the log-aware analysis; general expressions that can be applied to the expansion of any observable are given in \ref{app:bayes}.
We begin by reorganizing the chiral expansion in terms of
\begin{equation}
    \ell=\log\frac{\Mpi^2}{\Lb^2},\qquad \bar l_i(\Lb)=\bar l_i(\Mpi)+\ell=\bar l_i-\log\frac{\Lb^2}{(\Mpi^\text{phys})^2},
\end{equation}
i.e., Eq.~\eqref{eq:lbar} with $\Mpi^\text{phys}$ replaced by $\Lb$. Since $F_\pi$ does not depend on $\mu$, we can choose $\mu=\Lb$ in Eq.~\eqref{eq:fpi_corrections}, so that $L_\mu=\ell$, and obtain, through NNLO,
\begin{equation}
 \frac{F_\pi}{F}=1+q\big(\aleph_{4,0}+\aleph_{4,1}\ell\big)
 +q^2\big(\aleph_{6,0}+\aleph_{6,1}\ell+\aleph_{6,2}\ell^2\big)+\cdots,
\label{eq:logaware}
\end{equation}
where we again use the chiral-limit value $F$ as the reference scale and label the coefficients with respect to the chiral order instead of the power in $q$.
The coefficients of the rearranged expansion are
\begin{align}
\aleph_{4,0}
&=
\rho\,\bar l_4(\Lb),
\qquad
\aleph_{4,1}
=
-\rho,
\qquad
\aleph_{6,2}
=
-\frac{3}{4}\rho^2
\notag\\
\aleph_{6,0}
&=
\rho^2
\bigg[
r_F^r(\Lb)-\frac{13}{192}-\frac{\bar l_1(\Lb)+4\bar l_2(\Lb)-6\bar l_3(\Lb)\bar l_4(\Lb)}{12}
\bigg],
\notag\\
\aleph_{6,1}
&=
\rho^2
\bigg[
\frac{29}{12}+\frac{7\bar l_1(\Lb)+8\bar l_2(\Lb)}{6}-\bar l_4(\Lb)
\bigg],
\end{align}
with
\begin{equation}
\rho=\frac{\Lb^2}{(4\pi F)^2},
\end{equation}
so that $q\rho=\xi$, and $r_F^r(\Lb)$ follows from $r_F^r(770\MeV)=0$ by the RG running.
In contrast to the original form, the scale $\Lb$ now appears explicitly both in the polynomial terms and inside the chiral logarithms, while the renormalization scale no longer appears.
The coefficients thus coincide with those of the original expansion for $\mu=\Lb$, $\aleph_{n,m}=d_{i,m}(\mu=\Lb)$ with $n=2i+2$, e.g., $\aleph_{4,0}=d_{1,0}(\Lb)=\rho\,\bar l_4(\Lb)=16\pi^2\rho\, l_4^r(\Lb)$. The log-aware analysis therefore amounts to imposing naturalness on the renormalized coefficients at the scale $\mu=\Lb$. Note that this does not correspond to a fixed choice of renormalization scale: $\mu$ is tied to $\Lb$ and varies as the posterior for $\Lb$ is explored, in contrast to the fixed-log-scale variant mentioned below.  

Instead of assigning the naturalness assumption to the overall coefficient at a given order in the chiral expansion, the log-aware analysis applies the naturalness model to each coefficient $\aleph_{n,m}$ individually.
We collect these coefficients in the vector $\pmb{\aleph}=(\aleph_{4,0},\aleph_{4,1},\aleph_{6,0},\aleph_{6,1},\aleph_{6,2},\ldots)^T$,
where the number of entries depends on the order at which the inference is performed, and model them as independent Gaussian random variables with a common variance $\bar{c}^2$,
\begin{equation}
    \pmb{\aleph} \sim \mathcal{N}(0,\bar{c}^2)~,
\end{equation}
with the hyperprior of Eq.~\eqref{eq:hyperprior} for $\bar{c}^2$ and a log-uniform prior for $\Lb$.
The leading-log (LL) coefficients, $\aleph_{4,1}$ and $\aleph_{6,2}$, are fixed by lower-order terms. We nevertheless include them as independent naturalness draws (``with LL''); excluding them changes the median of $\Lb$ for $F_\pi$ only marginally, from $0.908\GeV$ to $0.885\GeV$. 
Because the log and non-log terms are treated separately, the $\aleph_{n,m}$ are pion-mass independent, and the results of the log-aware analysis do not depend on any choice of pion masses.
The resulting posterior for $\Lb$, including the Jacobian of the transformation from the original coefficients to the $\aleph_{n,m}$, is derived in \ref{app:bayes}.

Alternatively, one can keep $\mu$ fixed in the arguments of the logarithms, while still treating log and non-log terms separately as in the log-aware approach (referred to as the ``fixed-log-scale'' variant in the following).
While the expansion coefficients remain independent of the pion mass, they are explicitly $\mu$ dependent in this approach, and so will be the inferred scale $\Lb$. We use this variant as a check of the robustness of the log-aware results.

\begin{table}[t]
\caption{NNLO posterior medians and 68\% highest-posterior-density (HPD) intervals for $\Lb$
in GeV for $F_\pi$ from the pointwise analysis using different pion mass pairs and from the log-aware analysis.}
\label{tab:PointVsLog}\vspace{0.2cm}
\renewcommand{\arraystretch}{1.3}
\centering
\scalebox{0.9}{
\begin{tabular}{lrr}
\toprule
Analysis & Median & 68\% HPD\\
\midrule
Pointwise $\{30,530\}$\MeV & 0.660 & $[0.494,0.794]$\\
Pointwise $\{140,530\}$\MeV & 0.905 & $[0.692,1.087]$\\
Pointwise $\{140,1000\}$\MeV  & 0.934 & $[0.750,1.108]$\\
Log-aware with LL & 0.908 & $[0.806,1.008]$\\
\bottomrule
\end{tabular}}
\end{table}

\section{Results and convergence}
\label{sec:results}

Both the pointwise and the log-aware approaches are based on finding values of the scale $\Lb$ for which the coefficients of the chiral expansion are of natural size.
In the pointwise approach, in which there is only one (mass-dependent) coefficient at each order, there is a direct connection between naturalness and the breakdown scale: as long as the coefficients in Eq.~\eqref{eq:EFTexpand} are all roughly of the same, natural size, the convergence of the series is governed by the size of $Q$, and at $Q=1$ all contributions are expected to be of approximately the same size.
In the log-aware approach, the connection is less clear. With two or more coefficients at a given chiral order, cancellations or enhancements between these coefficients can occur, potentially obscuring the value of the soft scale at which different chiral orders become of the same size.
To see whether the extracted naturalness scales are consistent with the interpretation of a breakdown scale, all contributions at a given chiral order are collected into an effective, mass-dependent coefficient, analogous to the expansion of Eq.~\eqref{eq:EFTexpand},
\begin{align}\label{eq:XChiralExpand}
    \frac{X}{X_\text{LO}} & = 1 + c_1^\text{eff}(\Mpi,\Lb) q + c_2^\text{eff}(\Mpi,\Lb) q^2 + \cdots \notag\\
    & = 1+\Delta_\text{NLO}+\Delta_\text{NNLO}+\cdots\,,
\end{align}
where $\Delta_\text{NLO/NNLO}$ are the complete NLO/NNLO contributions to the observable $X$ after normalization to  $X_\text{LO}$.

\begin{figure}[t]
\centering
\includegraphics[width=\columnwidth,clip=true]{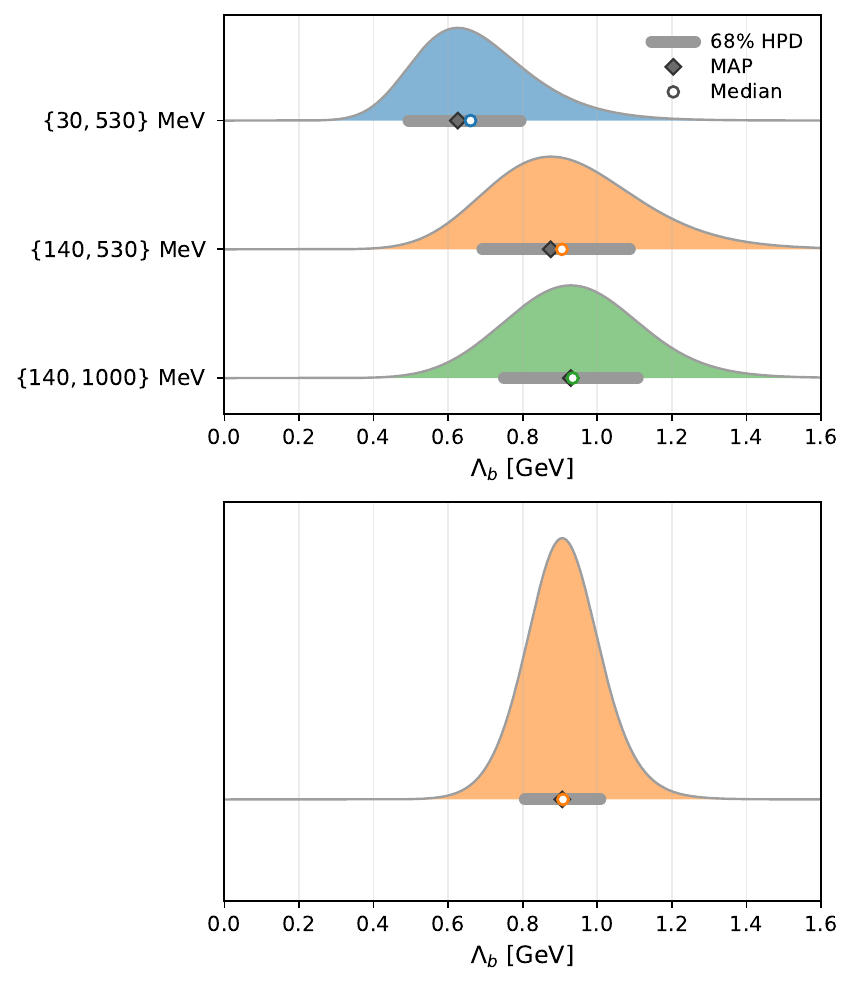}
\vspace{-0.7cm}
\caption{NNLO posterior distributions for $\Lb$ inferred from
$F_\pi$. The top panel shows the pointwise analyses using the
pion-mass pairs $\{30,530\}\MeV$, $\{140,530\}\MeV$, and
$\{140,1000\}\MeV$. The bottom panel shows the log-aware analysis
with the LL coefficients included as independent naturalness
draws. Each posterior distribution is normalized to its own
maximum. 
}
\label{fig:pointwise-logaware}
\end{figure}

\subsection{Pion decay constant}
\label{ssec:fpi}

Table~\ref{tab:PointVsLog} collects the main NNLO comparison for $F_\pi$ between the pointwise and the log-aware approaches with the LL coefficients included in the naturalness assumption. The corresponding posterior shapes are shown in Fig.~\ref{fig:pointwise-logaware}, with the results from the pointwise approach with different pion masses in the top panel and the log-aware approach in the bottom panel.
As a check of the log-aware results, we also perform a fixed-log-scale analysis.
As expected, the results exhibit noticeable dependence on the choice of $\mu$, with medians of $1.07\GeV$, $1.22\GeV$, and $0.89\GeV$ for $\mu=770\MeV$, $500\MeV$, and $140\MeV$, respectively, so that the commonly used choice $\mu=770\MeV$ results in a larger scale than the log-aware approach.

For the convergence of the chiral expansion, Eq.~\eqref{eq:XChiralExpand}, we evaluate the effective coefficients at the median extracted scale, denoted by $\tLb$ in the following.
Table~\ref{tab:Fpiceff} shows the size of these effective coefficients for pion masses in the range $0.140\GeV \le \Mpi \le \tLb$, with $\tLb=0.908\GeV$.
The coefficients are larger at small pion masses, as expected from their dependence on the chiral logarithms, but remain compatible with the underlying naturalness assumption in the entire pion-mass range.
The table also shows the NLO and NNLO contributions, $\Delta_\text{NLO}$ and $\Delta_\text{NNLO}$, respectively, for the same pion-mass range. The NNLO contribution remains suppressed relative to the NLO one for most of the pion masses, only becoming comparable in size for $\Mpi\simeq \tLb$.
Both $\Delta_\text{NLO}$ and $\Delta_\text{NNLO}$ remain less than 1 throughout the considered pion-mass range.
These results indicate that for $F_\pi$ the inferred naturalness scale is consistent with an interpretation as a breakdown scale.

\begin{table}[t]
\caption{Effective coefficients and order-by-order contributions for $F_\pi$ in the log-aware approach for fixed $\tLb = 0.908\GeV$ in the pion-mass range $0.140\GeV\le \Mpi \le \tLb$. Pion masses in \GeV, and $c_i^\text{eff} = c_i^\text{eff}(\Mpi,\tLb)$. 
}\vspace{0.2cm}
\renewcommand{\arraystretch}{1.3}
\label{tab:Fpiceff}
\centering
\scalebox{0.9}{
\begin{tabular}{lrrrrr}
\toprule
$\Mpi$ 
& $c_1^\text{eff}$
& $c_2^\text{eff}$
& $\Delta_\text{NLO}$  & $\Delta_\text{NNLO}$
& $\left|\frac{\Delta_\text{NNLO}}{\Delta_\text{NLO}}\right|$ \\
\midrule
$0.140$ & $2.992$ & $-1.448$ & $0.071$ & $-0.001$ & $0.012$ \\
$0.332$ & $1.780$ & $0.346$ & $0.238$ & $0.006$ & $0.026$ \\
$0.524$ & $1.140$ & $0.404$ & $0.380$ & $0.045$ & $0.118$ \\
$0.716$ & $0.701$ & $0.089$ & $0.436$ & $0.035$ &  $0.079$ \\
$0.908$ & $0.368$ & $-0.343$ &  $0.368$ & $-0.343$ & $0.932$ \\
\bottomrule
\end{tabular}}
\end{table}

\subsection{\texorpdfstring{$\pi\pi$}{pipi} scattering}
\label{sec:pipi_results}

We extend the application of the log-aware method to the $\pi\pi$ threshold and subthreshold parameters introduced in Sec.~\ref{ssec:pipi}.
We treat each quantity separately and do not consider any correlations among them.
Similarly to the case of $F_\pi$, in the pointwise approach the extracted scales $\Lb$ show dependence on the choice of pion masses used in the analysis.
In the following, we therefore focus on the results obtained in the log-aware approach.
Table~\ref{tab:convergence-median} shows the median inferred values of $\Lb$ for the scattering parameters with the associated 68\% HPD intervals for the two sets of NNLO LECs $r_i$ defined in Sec.~\ref{ssec:lecs}.
There are considerable differences in the inferred scales of different
quantities even at NNLO, with median values ranging from
$0.345\GeV$ for $b_0^0$ to $0.838\GeV$ for $c_0^0$ for Set~I and from $0.403\GeV$ for $b_0^0$ to $0.838\GeV$ for $c_0^0$ for Set~II.
The impact of the choice of LECs differs among observables: $c_0^0$ is unaffected, since the combination $27r_1+36r_2+48r_3+64r_5$ entering at NNLO vanishes for vector-meson saturation, Eq.~\eqref{eq:ri_res}; $s_A$ is almost unaffected. 
$\tLb$ for $a_0^0$, $a_1^1$, $b^0_0$ and $c_1^0$ increases by up to about $0.15\GeV$ from Set I to Set II, and $\tLb$ for $a_0^2$ and $b_0^2$ moves in the opposite direction.

\begin{table}[t]
\caption{
Log-aware posterior medians $\tLb$ and 68\% HPD intervals for $\pi\pi$ scattering observables, as well as contributions to the chiral expansion at $\Mpi=\tLb$ for the two sets of $r_i$.}
\label{tab:convergence-median}\vspace{0.2cm}
 \renewcommand{\arraystretch}{1.3}
\centering
\scalebox{0.9}{
\begin{tabular}{llrrrr}
\toprule
 & & $\tLb$ [GeV] & $\Delta_\text{NLO}$ & $\Delta_\text{NNLO}$ & $|\Delta_\text{NNLO}/\Delta_\text{NLO}|$\\
\midrule
\multirow{2}{*}{$a_0^0$} & Set I       & $0.632^{+0.093}_{-0.106}$ &$  0.525 $&$ -0.801 $&$ 1.526 $\\
& Set II &
$0.765^{+0.184}_{-0.218}$ &$ 0.293 $&$ 3.873 $&$ 13.230 $\\
\multirow{2}{*}{$a_1^1$} & Set I       &$ 0.391^{+0.057}_{-0.064} $&$  0.582 $&$  1.850 $&$ 3.181 $\\
& Set II&$ 
0.536^{+0.060}_{-0.059} $&$ 1.092 $&$ 0.101 $&$ 0.092 $\\
\multirow{2}{*}{$a_0^2$} & Set I       &$ 0.624^{+0.089}_{-0.108} $&$  0.248 $&$ -1.552 $&$ 6.268 $\\
& Set II&$ 
0.520^{+0.075}_{-0.087} $&$ -0.122 $&$ 1.915 $&$ 15.726 $\\
\multirow{2}{*}{$b_0^0$} & Set I       &$ 0.345^{+0.044}_{-0.050} $&$  0.302 $&$ -1.361 $&$ 4.501 $\\
& Set II&$
0.403^{+0.053}_{-0.060} $&$ 0.124 $&$ -1.205 $&$ 9.693 $\\
\multirow{2}{*}{$b_0^2$} & Set I       &$ 0.541^{+0.078}_{-0.092} $&$ -0.449 $&$ -1.635 $&$ 3.644 $\\
& Set II&$ 
0.471^{+0.076}_{-0.089}  $&$ -0.618 $&$ 2.355 $&$ 3.809 $\\
\multirow{2}{*}{$s_A$} & Set I         &$ 0.505^{+0.066}_{-0.067} $&$  1.130 $&$ -0.231 $&$ 0.204 $\\
& Set II&$ 
0.508^{+0.066}_{-0.069} $&$ 1.151 $&$ -0.049 $&$ 0.043 $\\
\multirow{2}{*}{$c_0^0$} & Set I   &$ 0.838^{+0.115}_{-0.110} $&$ -1.325 $&$ -0.617 $&$ 0.465 $\\
& Set II&$
0.838^{+0.115}_{-0.110} $&$ -1.325 $&$ -0.617 $&$ 0.465 $\\
\multirow{2}{*}{$c_1^0$} & Set I   &$ 0.612^{+0.085}_{-0.099} $&$  0.202 $&$  1.820 $&$ 9.012 $\\
& Set II&$
0.773^{+0.091}_{-0.091} $&$ 0.323 $&$ 1.295 $&$ 4.012 $\\
\bottomrule
\end{tabular}}
\end{table}

The connection between the inferred naturalness scales and the convergence of the chiral expansions of the various scattering quantities is analyzed in analogy to the case of $F_\pi$, see Eq.~\eqref{eq:XChiralExpand}.
Table~\ref{tab:convergence-median} shows the NLO and NNLO corrections evaluated at $\Mpi$ corresponding to the median inferred value $\tLb$ for each observable.
The individual contributions $|\Delta_\text{NLO}|$ and $|\Delta_\text{NNLO}|$ for some cases are larger than 1, but still consistent with the underlying naturalness assumption. 
While the ratio $|\Delta_\text{NNLO}/\Delta_\text{NLO}|$ becomes markedly larger than 1 for several observables, this is the result of an interplay of relatively small NLO contributions even at $\Mpi=\tLb$ and relatively large NNLO contributions. For instance, $\Delta_\text{NLO}$ of $a_0^2$ changes sign at $\Mpi\simeq 0.56\GeV$, between the values of $\tLb$ for the two sets. 

\begin{figure*}[t]
  \centering
  \includegraphics[width=0.32\textwidth]{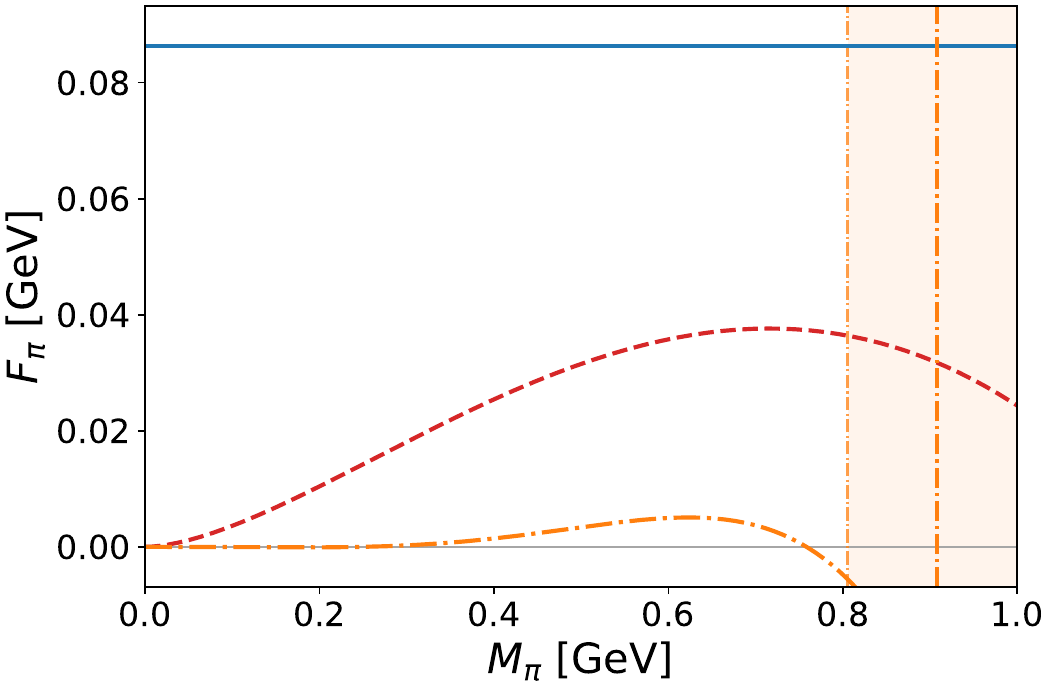} \hfill
  \includegraphics[width=0.32\textwidth]{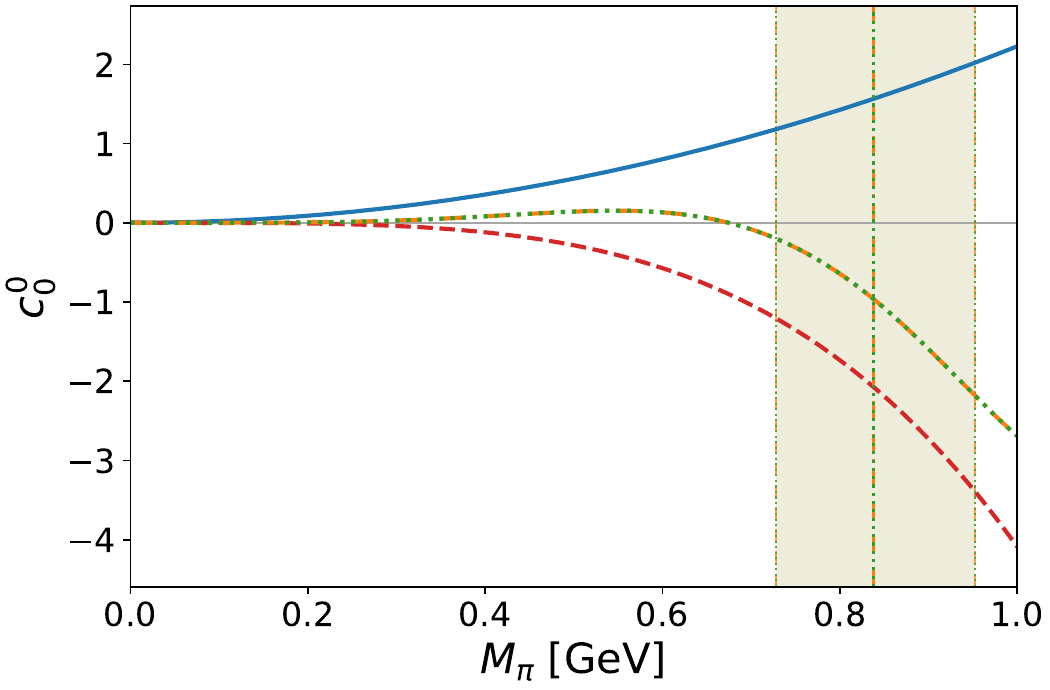} \hfill
  \includegraphics[width=0.32\textwidth]{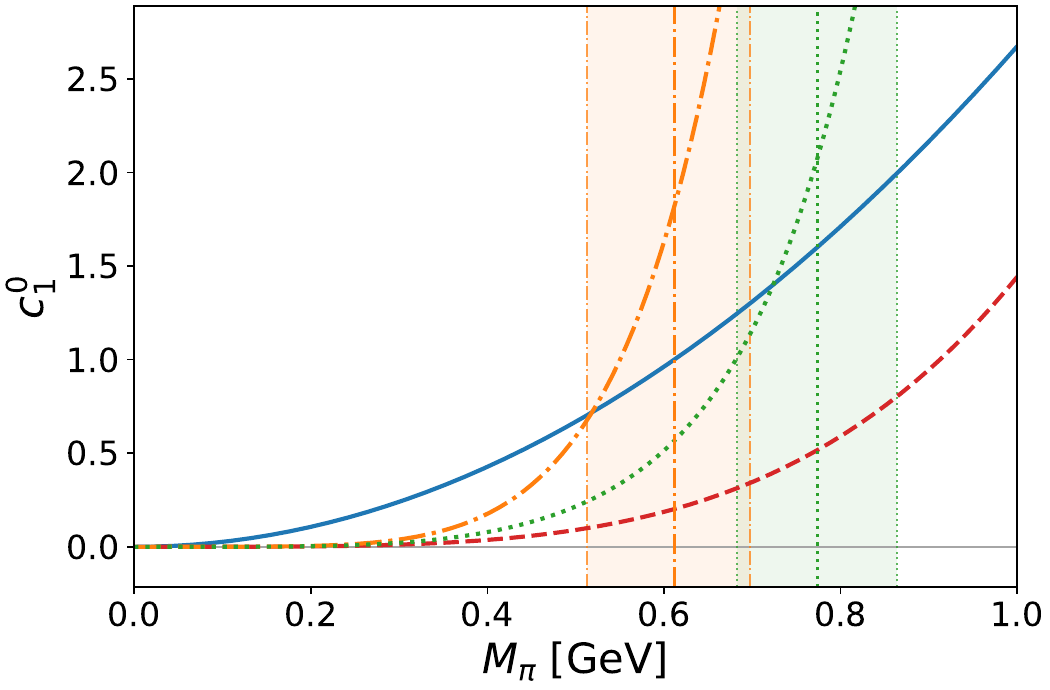}

  \medskip

  \includegraphics[width=0.32\textwidth]{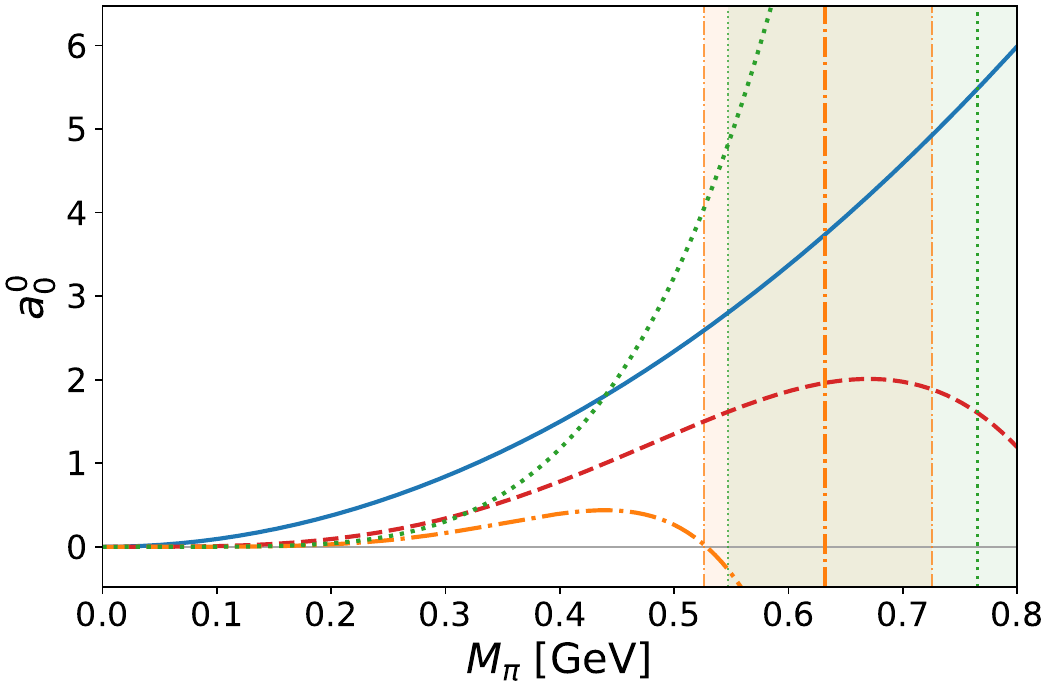} \hfill
  \includegraphics[width=0.32\textwidth]{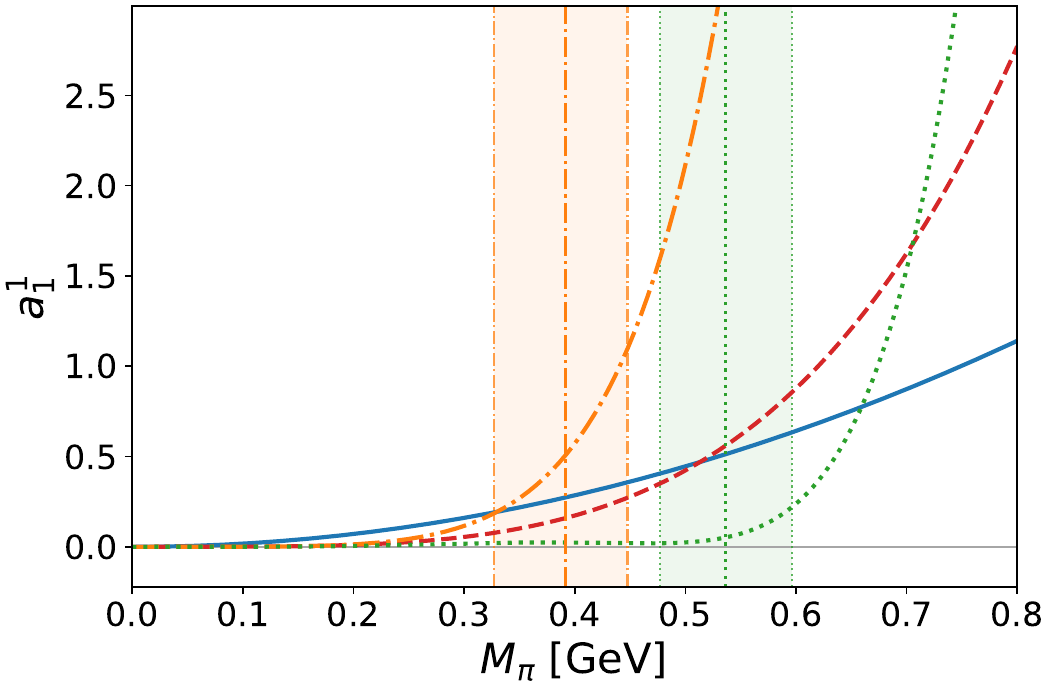} \hfill
  \includegraphics[width=0.32\textwidth]{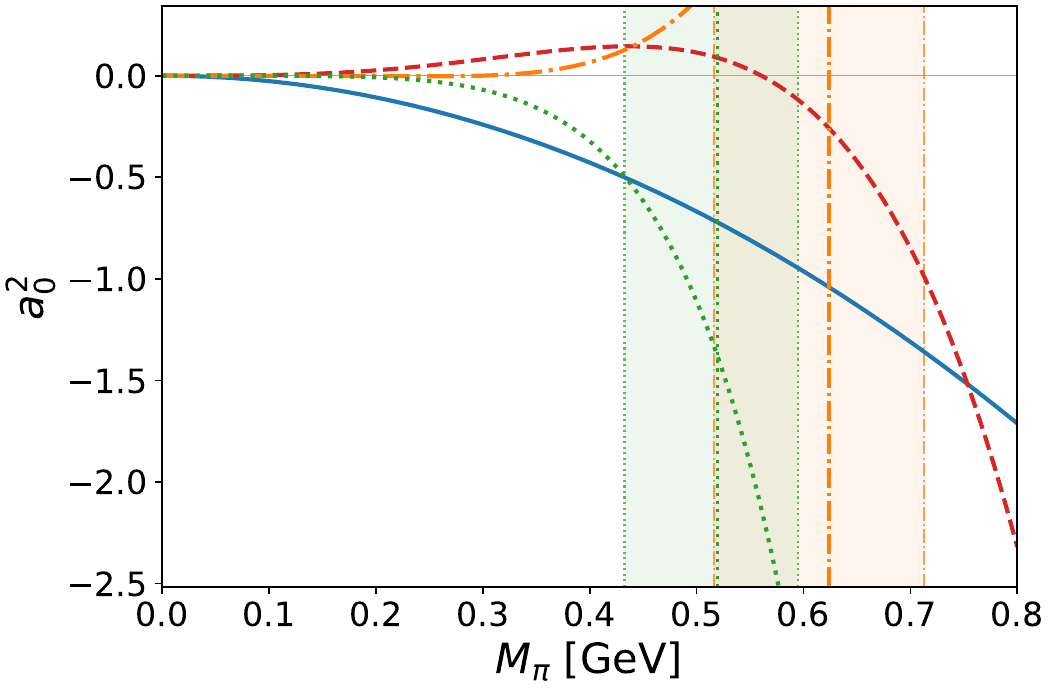}

  \medskip
  
  \includegraphics[width=0.32\textwidth]{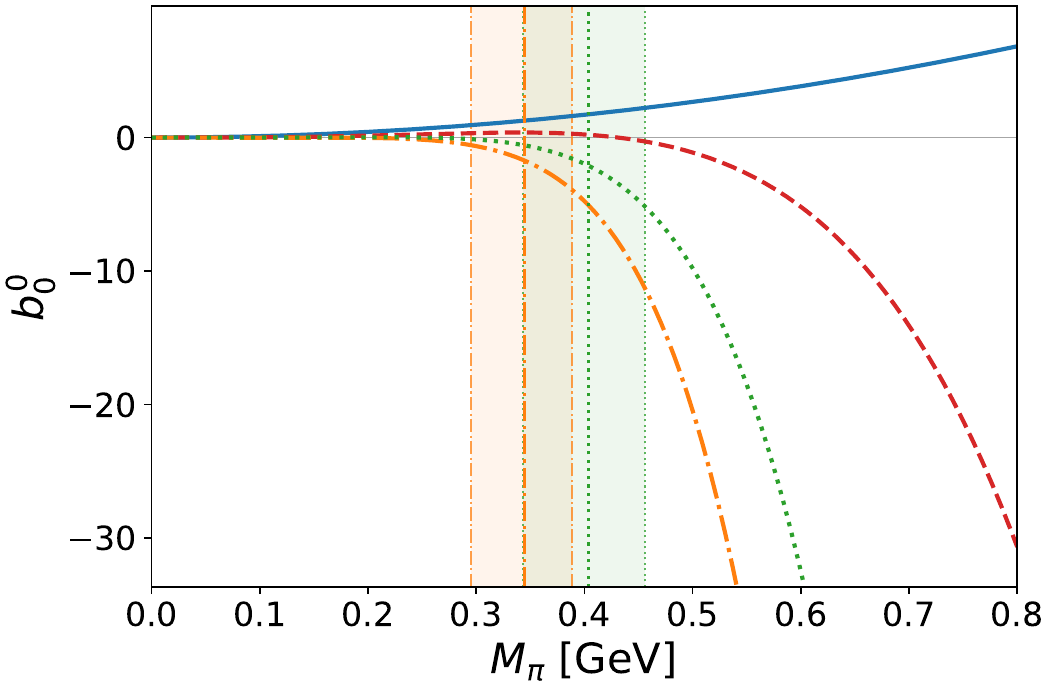} \hfill
  \includegraphics[width=0.32\textwidth]{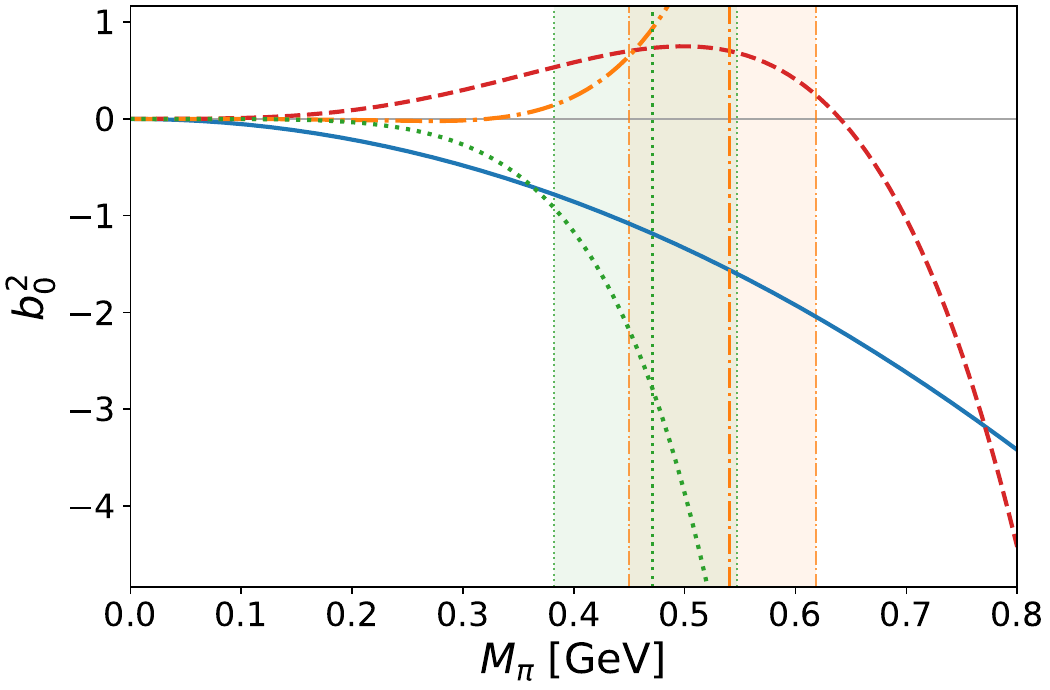} \hfill
  \includegraphics[width=0.32\textwidth]{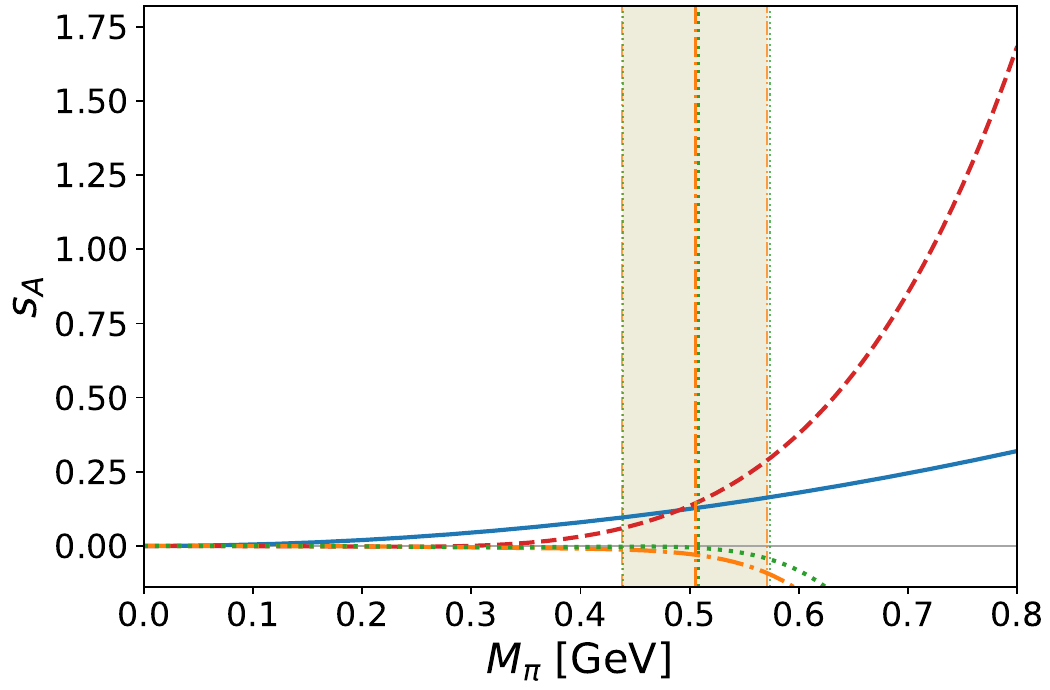}
  
\medskip
\includegraphics[width=0.6\textwidth]{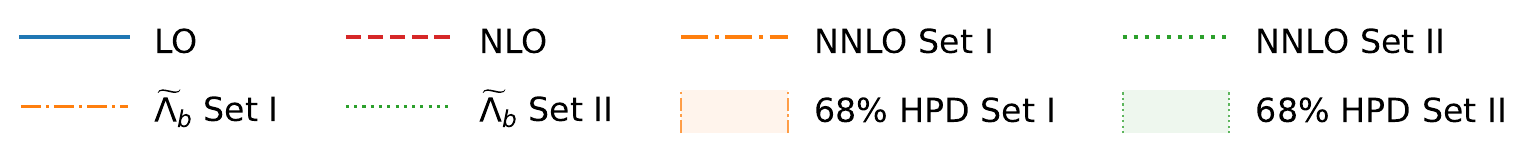}

  \caption{LO, NLO, and NNLO contributions as functions of
    $\Mpi$, with the NNLO contributions shown for both sets of $r_i$. Vertical lines mark the
    corresponding NNLO posterior medians $\tLb$ for each set, with the bands corresponding to the 68\% HPD intervals. $F_\pi$ is independent of the Set I and II LECs.}
  \label{fig:convergence}
\end{figure*}

These patterns are further illustrated in Fig.~\ref{fig:convergence}, which shows that, judged by the size of the individual corrections rather than by their ratio, the inferred naturalness scale is mostly reasonably close to the point at which the order-by-order convergence breaks down. The most notable exception concerns $a_0^0$ for Set~II, for which $\tLb$ is much larger than the breakdown scale one would have assigned based on the hierarchy of subsequent orders, with $\Delta_\text{NNLO}\simeq 3.9$ at $\Mpi=\tLb$. In this case, one finds that the non-log, single-log, and double-log terms, individually, are indeed of natural size, but conspire in the same direction, so that the overall convergence breaks down significantly earlier than expected from $\tLb$. Finally, the convergence should improve for the subthreshold compared to the threshold parameters given the increased distance from unitarity cuts, which  holds true for $c_0^0$, while the range of convergence for $c_1^0$ tends to be lower, depending on the NNLO LECs. For the Adler zero $s_A$, the NNLO correction remains small, but the NLO one already reaches the size of the LO at $\Mpi\simeq\tLb\simeq 0.5\GeV$ (for $c_0^0$ only at $\Mpi\simeq 0.7\GeV$), so that its convergence only improves marginally compared to the threshold parameters. In general, the convergence properties are related to the size of the coefficients of the chiral logarithms, which for $F_\pi$ are ${\mathcal O}(1)$, but can become ${\mathcal O}(10)$ for the threshold parameters. The exact range of convergence, however, depends on the LECs and the relative signs of the different contributions.

\section{Lattice-QCD input for the \texorpdfstring{$\Order(p^6)$}{O(p6)} LECs}
\label{sec:lattice}

The results of Sec.~\ref{sec:pipi_results} show that the inferred scales depend on the resonance-saturation estimates of the $r_i$, which also enter the matching of Roy-equation analyses to ChPT~\cite{Colangelo:2001df}. At fixed pion mass, the $\Order(p^6)$ amplitude can be expressed in terms of six coefficients $b_{1\text{--}6}$, which parametrize its polynomial part~\cite{Bijnens:1995yn,Bijnens:1997vq}, so that shifts in $\bar l_1$ and $\bar l_2$ can be compensated by the $r_i$ without changing the $b_i$. The $\bar l_i$ extracted from the matching therefore depend on the input for $r_{1\text{--}4}$: repeating the matching of Ref.~\cite{Colangelo:2001df} with the same Roy-equation input, but with the $r_i$ of Table~\ref{tab:ri}, we obtain $\bar l_1=-0.48$, $\bar l_2=4.35$ for Set~II and $\bar l_1=0.48$, $\bar l_2=3.90$ for Set~I, compared with $\bar l_1=-0.4(6)$, $\bar l_2=4.3(1)$~\cite{Colangelo:2001df}, where the slight difference to Set II originates from including scalar and kaon-/$\eta$-loop contributions~\cite{Bijnens:1997vq}. 

Since the $\bar l_i$ enter at $\Order(\xi)$ and in the coefficients of the chiral logarithms, whereas the $r_i$ enter only at $\Order(\xi^2)$ and in polynomials, the compensation holds only at the pion mass at which it is imposed, so that lattice-QCD calculations at unphysical pion masses can lift the degeneracy. Once the $b_i$ are fixed at the physical point, the change of a threshold parameter at another pion mass is dominated by the NLO term, proportional to $\xi(\Mpi)-\xi(\Mpi^\text{phys})$ times the combination of $\bar l_1$ and $\bar l_2$ that appears at NLO. These combinations differ among the observables of Sec.~\ref{ssec:pipi}: $a_0^0$ and $a_0^2$ depend on $\bar l_1+2\bar l_2$, $b_0^0$ on $2\bar l_1+3\bar l_2$, $b_0^2$ on $\bar l_1+3\bar l_2$, and $a_1^1$ on $\bar l_2-\bar l_1$, see \ref{app:threshold-equations}, so that the $I=0$ and $I=2$ channels probe almost the same combinations. Given that the most precise lattice-QCD calculations are available for $I=2$~\cite{NPLQCD:2011htk,ETM:2015bzg,Horz:2019rrn,Mai:2019pqr,Fischer:2020jzp,Blanton:2021llb,RBC:2023xqv,Dawid:2025doq}, due to the simplicity of the channel with a weak, repulsive interaction and no resonances, we focus on $I=2$ and $I=1$ in the following. 
The uncertainty of the current matching, obtained by varying the $r_i$ as in Ref.~\cite{Colangelo:2001df} and including the resulting correlation between $\bar l_1$ and $\bar l_2$, amounts to about $0.5$ for $\bar l_1+2\bar l_2$ and $0.6$ for $\bar l_2-\bar l_1$. From Set~II to Set~I, $\bar l_1$ increases by $0.96$ and $\bar l_2$ decreases by $0.45$, so that $\bar l_1+2\bar l_2$ changes by only $0.06$, whereas $\bar l_2-\bar l_1$ changes by $-1.41$. These numbers define the benchmarks at which level additional lattice-QCD input would have an impact. 

For $I=2$, imposing  the requirement $|\Delta_\text{NNLO}/\Delta_\text{NLO}|\lesssim 0.5$ suggests that despite the absence of resonances the discussion should be restricted to $\Mpi\lesssim 250\MeV$, see Fig.~\ref{fig:convergence}. 
 In this range, the NNLO corrections modify the combination probed by $a_0^2$ only slightly, to $\bar l_1+\kappa\,\bar l_2$ with $\kappa=2.1$--$2.2$, so that data at several pion masses constrain the same combination. Moreover, $b_0^2$ adds little independent information, and also the $D$-wave probes a similar combination, $\bar l_1+\bar l_2$, in addition to a sign change at $\Mpi\simeq (155\text{--}200)\MeV$ that implies delicate cancellations. Accordingly, the $I=2$ channel essentially constrains $\bar l_1+2\bar l_2$: a unit shift in this combination, compensated by the $r_i$ at the physical point, changes $a_0^2$ by about $6\%$ at $200\MeV$ and $18\%$ at $250\MeV$. As shown in Table~\ref{tab:lattice}, determinations of $a_0^2$ at these pion masses with a precision of $5\%$ would improve on the current matching by a factor of about two, and with $2\%$ by a factor of about four. In contrast, the $I=2$ channel by itself is insensitive to the difference between Set~I and Set~II, since, as discussed above, in $\bar l_1+2\bar l_2$ the changes happen to almost compensate.
 
 In addition, the $r_i$ of the two sets differ beyond the compensation of the shift in the $\bar l_i$. Since $a_0^2$ depends on the $r_i$ only through the combination $r_1+16r_4$, see \ref{app:threshold-equations}, which equals $-4\gamma$ in Eq.~\eqref{eq:ri_res} and is thus driven by $f_\chi^2$, the two sets predict values of $a_0^2$ that differ by about $1\%$ at the physical point, within the uncertainty of the Roy-equation input, but by $6\%$ at $200\MeV$ and $15\%$ at $250\MeV$. Hence, $a_0^2$ also provides a direct test of the resonance-saturation input: already a $5\%$ determination at $250\MeV$ would distinguish the two sets at the $3\sigma$ level. This test becomes more stringent at larger pion masses, but at the expense of reduced control over the chiral expansion.

\begin{table}[t]
\caption{Projected constraints from lattice-QCD determinations of $a_0^2$ and $a_1^1$ at $\Mpi=200$ and $250\MeV$, with relative precision $\epsilon$, on the combinations of $\bar l_1$ and $\bar l_2$ ($b_i$ fixed at the physical point, starting from Set~II, linearized, truncation uncertainty not included). The last column gives the significance with which the shift between Sets~I and~II in $\bar l_1$, $\bar l_2$ would be resolved; the difference in their predictions for $a_0^2$ through $r_1+16r_4$, see main text, is not included. The current matching corresponds to $\sigma(\bar l_1+2\bar l_2)\simeq0.5$ and $\sigma(\bar l_2-\bar l_1)\simeq0.6$, with $\sigma$ referring to the respective error.}
\label{tab:lattice}\vspace{0.2cm}
\renewcommand{\arraystretch}{1.3}
\centering
\scalebox{0.9}{
\begin{tabular}{llrrrr}
\toprule
$\epsilon(a_0^2)$ & $\epsilon(a_1^1)$ & $\sigma(\bar l_1+2\bar l_2)$ & $\sigma(\bar l_2-\bar l_1)$ & $\sigma(\bar l_1)$, $\sigma(\bar l_2)$ & I vs.\ II\\
\midrule
$5\%$ & -- & $0.27$ & -- & -- & $\simeq 0$ \\
$2\%$ & -- & $0.11$ & -- & -- & $\simeq 0$ \\
$2\%$ & $10\%$ & $0.11$ & $0.88$ & $0.60$, $0.28$ & $1.6\sigma$ \\
$2\%$ & $5\%$ & $0.10$ & $0.44$ & $0.31$, $0.14$ & $3.2\sigma$ \\
$2\%$ & $3\%$ & $0.10$ & $0.27$ & $0.19$, $0.09$ & $5.3\sigma$ \\
\bottomrule
\end{tabular}}
\end{table}

 For $a_1^1$, the NNLO corrections modify its NLO combination, $\bar l_2-\bar l_1$, more strongly, to approximately $2\bar l_2-\bar l_1$ at $\Mpi=(200\text{--}250)\MeV$, which remains sufficiently distinct from the $a_0^2$ direction. Combined with $a_0^2$ at $2\%$, a determination of $a_1^1$ at these pion masses with a precision of $5\%$ would distinguish Sets~I and~II at the $3\sigma$ level, and a precision of $3\%$ would improve on the current matching for $\bar l_2-\bar l_1$ by a factor of about two, see Table~\ref{tab:lattice}. This is more demanding than for $I=2$, since $a_1^1$ needs to be extracted from a channel dominated by the $\rho$ resonance, which requires a parameterization of the resonance and an extrapolation to threshold. Indeed, $I=1$ lattice calculations, including at $\Mpi\simeq (200\text{--}250)\MeV$ and $M_\pi^\text{phys}$~\cite{Wilson:2015dqa,Andersen:2018mau,Mai:2019pqr,Rodas:2023gma,Boyle:2024grr,Jevsenak:2026pok}, have so far focused on the $\rho$ resonance parameters rather than on $a_1^1$. Moreover, its chiral expansion converges more slowly: for Set~I, $|\Delta_\text{NNLO}/\Delta_\text{NLO}|$ of $a_1^1$ exceeds $1$ already at $\Mpi=200\MeV$, compared with about $0.5$ for Set~II, see Fig.~\ref{fig:convergence}.

These estimates assume the $b_i$ to be fixed by the Roy-equation input and neglect the truncation uncertainty, so that the quoted precisions are necessary rather than sufficient. If the amplitude is expressed in terms of $F_\pi$ instead of $F$, see Sec.~\ref{ssec:conventions}, the effects of such compensated shifts decrease by almost a factor of two, a difference that is formally of $\Order(p^8)$. Present calculations reach $(2$--$3)\%$ for $a_0^2$, but mostly at $\Mpi\gtrsim 280\MeV$, and $(6$--$11)\%$ at $200\MeV$ and at the physical point~\cite{Blanton:2021llb,Dawid:2025doq}. In summary, determinations of $a_0^2$ at the $(2$--$5)\%$ level and of $a_1^1$ at the level of $5\%$ for $\Mpi\simeq (200\text{--}250)\MeV$ would provide sensitivity to both combinations of $\Order(p^6)$ LECs that are currently estimated from resonance saturation.

\section{Summary}
\label{sec:summary}

In this Letter, we developed a new method to infer the naturalness scale in ChPT calculations that contain logarithmic dependence on the pion mass, the small scale of the EFT expansion. Using this approach, we considered the chiral expansions for the pion decay constant $F_\pi$ and $\pi\pi$ scattering parameters.

We find substantial observable dependence in the inferred naturalness scales ranging from approximately $350\MeV$ to $900\MeV$. For many cases the inferred naturalness scale is of the same order as the pion mass at which all terms in the chiral expansion are comparable in size. However, the parameter $a_0^0$ shows that the inferred naturalness scale does not have to correspond to the pion mass at which the order-by-order hierarchy breaks down, in case different contributions that individually are of natural size conspire in the same direction. 

The expectation of a better convergence in the subthreshold region is largely borne out, with subthreshold parameters displaying, in general, better convergence properties than threshold parameters, but results differ on a case-by-case basis and depend on the numerical values of the NNLO LECs.
We studied the sensitivity by comparing results for two different sets of resonance-saturation estimates. 

To obtain better constraints on these LECs, additional input from lattice QCD is necessary, since  
the $\pi\pi$ parameters depend on two linear combinations that the matching to Roy-equation analyses cannot determine. Lattice-QCD calculations at $\Mpi\simeq (200\text{--}250)\MeV$, within the range where the chiral expansion remains under control, could constrain them: $a_0^2$ at the $(2$--$5)\%$ level would fix $\bar l_1+2\bar l_2$ more precisely than the current matching, while a second, independent combination requires the $I=1$ $P$-wave, with $a_1^1$ at the level of $5\%$; the $I=0$ channel probes almost the same combinations as $I=2$.

\section*{Acknowledgments}
We thank Gilberto Colangelo and Bastian Kubis for useful discussions. 
We acknowledge use of the Python package `gsum'~\cite{Melendez:2019izc} as well as OpenAI's Codex (GPT-5 and GPT-6) for generating Python code for the log-aware analysis.
This work was supported by the US Department of Energy under Contract Nos.\ DE-AC05-00OR22725, the U.S.~Department of Energy, Office of Science, Office of Nuclear Physics (Award No.~DE-SC0019647), the INT's U.S.\ Department of Energy grant No.~DE-FG02-00ER41132, by the National Science Foundation (Grant No.~PHY-2412612), by the Ram\'on y Cajal program (RYC2019-027605-I) of the Spanish MINECO, by the MCIN/AEI/10.13039/501100011033 (grants PID2022-136510NB-C31 and CNS2025-166481), and by the Swiss National Science Foundation (Project No.\ TMCG-2\_213690).
    M.H.\ and M.R.S.\  
thank the Institute for Nuclear Theory at the University of Washington
for its kind hospitality and support during the program
``Nuclear Hamiltonians for Advancing Nuclear Physics and Beyond (26-1),'' during which this project was initiated.

\appendix

\section{\texorpdfstring{$\pi\pi$}{pipi} scattering amplitude parameters}
\label{app:threshold-equations}

The $\pi\pi$ scattering amplitude can be decomposed into partial-wave
amplitudes $t_\ell^I(s)$, where $I$ denotes the total isospin of the
two-pion state, $\ell$ the orbital angular momentum, and $s$ the squared center-of-mass energy. For even/odd isospin only even/odd partial waves contribute. 
The normalization of the isospin amplitudes is chosen such that
\begin{equation}
T^I(s,t,u)
=
32\pi
\sum_{\ell=0}^{\infty}
(2\ell+1)
P_\ell(\cos\theta)\,
t_\ell^I(s),
\end{equation}
where $\theta$ is the center-of-mass scattering angle.

The pion center-of-mass momentum $q_\text{CM}$ is made dimensionless according to
\begin{equation}
\hat q^2
\equiv
\frac{q_\text{CM}^2}{M_\pi^2}
=
\frac{s}{4M_\pi^2}-1.
\end{equation}
Near threshold, the
real part of the partial-wave amplitude is expanded as in Eq.~\eqref{threshold_expansion}.
With this convention, the threshold parameters $a_\ell^I$ and
$b_\ell^I$ are dimensionless as well. 

We also consider subthreshold parameters obtained by expanding the
isospin amplitudes around the symmetric point
$s=t=u=4M_\pi^2/3$.
Introducing the dimensionless variables
\begin{equation}
\hat{s}=\frac{s}{M_\pi^2},
\qquad
\hat{t}=\frac{t}{M_\pi^2},
\qquad
\hat{u}=\frac{u}{M_\pi^2},
\end{equation}
we define the subthreshold coefficients $c_i^I$ through
\begin{equation}
\frac{T^I(s,t,u)}{32\pi}
=
c_0^I
+
c_1^I\left(\hat{s}-\frac{4}{3}\right)
+
c_2^I\left(\hat{t}-\hat{u}\right)
+
\mathcal{O}(\epsilon^2),
\label{eq:subthreshold-expansion}
\end{equation}
where $\epsilon$ collectively denotes deviations from the symmetric
point.

Crossing symmetry relates the parameters in the different isospin
channels,
\begin{equation}
c_0^{2}=\frac{2}{5}c_0^{0},
\qquad
c_1^{2}=-\frac{1}{2}c_1^{0},
\qquad
c_2^{1}=\frac{1}{2}c_1^{0}.
\end{equation}
Only two of these subthreshold parameters are independent, and we
therefore use $c_0^{0}$ and $c_1^{0}$ in the naturalness-scale
analysis.
Finally, $s_A$ denotes the position of the Adler zero of the $I=0$
$S$-wave below threshold, $t_0^0(s_A)=0$, which at LO lies at
$s_A=M_\pi^2/2$.

In the following, $X^{(n)}$ denotes the $\Order(p^n)$ contribution to
the observable $X$, with $\bar l_i\equiv\bar l_i(\Mpi)$ and
$L_\mu$ defined as in Sec.~\ref{ssec:conventions}. The expressions used in this work were extracted from Refs.~\cite{Niehus:2020gmf,Heuser:2024biq}. For the scattering lengths and effective ranges they read:
\begin{align}
a_0^{0,(2)}
&=\frac{7\pi}{2}\xi,\qquad 
a_0^{0,(4)}
=\frac{5\pi}{12}\xi^2
\Big[21+8\bar l_1+16\bar l_2-3\bar l_3\Big],
\\
a_0^{0,(6)}
&=\frac{\pi}{576}\xi^3
\bigg[
31107+72\Big(202\bar l_1
+328\bar l_2+\bar l_3-10\bar l_3^2\Big)\notag\\
&-2240\pi^2+73728\pi^4\Big(5r_1^r+12r_2^r+48r_3^r+32r_4^r+192r_5^r\Big)\notag\\
&-4032r_F^r+240\Big(33-34\bar l_1-112\bar l_2
+24\bar l_3\Big)L_\mu
-14640L_\mu^2
\Big],\notag\\
a_0^{2,(2)}
&=-\pi\xi,\qquad 
a_0^{2,(4)}
=\frac{\pi}{6}\xi^2\Big[3+8\bar l_1+16\bar l_2-3\bar l_3\Big],
\\
a_0^{2,(6)}
&=-\frac{\pi}{288}\xi^3\bigg[
2973+72\Big(6\bar l_1+24\bar l_2+7\bar l_3+2\bar l_3^2\Big)-704\pi^2
\nonumber\\
&
-73728\pi^4\big(r_1^r+16r_4^r\big)-576r_F^r
\nonumber\\
&
-48\Big(45+38\bar l_1+32\bar l_2
-12\bar l_3\Big)L_\mu
-1392L_\mu^2
\Big],\notag\\
a_1^{1,(2)}
&=\frac{2\pi}{3}\xi,\qquad 
a_1^{1,(4)}
=-\frac{\pi}{54}\xi^2
\Big[65+48\bar l_1-48\bar l_2\Big],
\\
a_1^{1,(6)}
&=-\frac{\pi}{1296}\xi^3
\bigg[
4767+8\Big(422\bar l_1+1168\bar l_2
-531\bar l_3\Big)-2576\pi^2\notag\\
&-221184\pi^4\Big(r_2^r+8r_4^r+16r_6^r\Big)
+1728r_F^r\notag\\
&+16\Big(853+846\bar l_1+1284\bar l_2
-135\bar l_3\Big)L_\mu
+15960L_\mu^2
\bigg],\notag\\
b_0^{0,(2)}
&=4\pi\xi,\qquad 
b_0^{0,(4)}
=\frac{\pi}{3}\xi^2
\Big[-13+32\bar l_1+48\bar l_2\Big],
\\
b_0^{0,(6)}
&=\frac{\pi}{648}\xi^3
\bigg[
-39877+24\Big(1658\bar l_1+1232\bar l_2
+441\bar l_3\Big)\notag\\
&-29931\pi^2
+663552\pi^4\Big(r_2^r+12r_3^r+12r_4^r+72r_5^r-8r_6^r\Big)\notag\\
&-5184r_F^r
+48\Big(1399-582\bar l_1-1908\bar l_2
+135\bar l_3\Big)L_\mu\notag\\
&-56520L_\mu^2
\bigg],\notag\\
b_0^{2,(2)}
&=-2\pi\xi,\qquad 
b_0^{2,(4)}
=\frac{\pi}{6}\xi^2\Big[
-5+16\bar l_1+48\bar l_2\Big],
\\
b_0^{2,(6)}
&=\frac{\pi}{1296}\xi^3
\bigg[
-54275+24\Big(166\bar l_1-752\bar l_2
-387\bar l_3\Big)\notag\\
&+15216\pi^2
-663552\pi^4\Big(r_2^r-24r_4^r+16r_6^r\Big)
+5184r_F^r\notag\\
&+48\Big(845+654\bar l_1+900\bar l_2
-135\bar l_3\Big)L_\mu
+34056L_\mu^2
\bigg].\notag
\end{align}
For the subthreshold parameters we obtain
\begin{align}
c_0^{0,(2)}
&=\frac{5\pi}{6}\xi,\\
c_0^{0,(4)}
&=\frac{5\pi}{108}\xi^2
\Big[33+8\bar l_1+16\bar l_2-27\bar l_3-66\sigma\Big],\notag
\\
c_0^{0,(6)}
&=\frac{5\pi}{139968}\xi^3
\bigg[
1033493-922464\sigma+4032\sigma^2+136080\sigma^3\notag\\
&+144\bar l_1\big(707-912\sigma\big)+192\bar l_2\big(1309-1560\sigma\big)\notag\\
&-5832\bar l_3\big(17-44\sigma\big)-34992\bar l_3^2+24\pi^2\big(818-2835\sigma\big)\notag\\
&+663552\pi^4\Big(27r_1^r+36r_2^r+48r_3^r+64r_5^r\Big)-46656r_F^r\notag\\
&-48\Big(7805+3366\bar l_1+5136\bar l_2-1944\bar l_3\Big)L_\mu-157392L_\mu^2
\bigg],\notag\\
c_1^{0,(2)}
&=\pi\xi,
\\
c_1^{0,(4)}
&=\frac{\pi}{144}\xi^2
\Big[2-64\bar l_1+64\bar l_2-243\sigma\Big],\notag
\\
c_1^{0,(6)}
&=\frac{\pi}{7776}\xi^3
\bigg[
20237-40530\sigma+47304\sigma^2-1134\sigma^3\notag\\
&+48\bar l_1\big(413-339\sigma\big)
+96\bar l_2\big(226-303\sigma\big)\notag\\
&+486\bar l_3\big(2+65\sigma\big)
+3\pi^2\big(134+189\sigma\big)\notag\\
&+663552\pi^4\Big(3r_2^r+8r_3^r+16r_5^r\Big)-15552r_F^r\notag\\
&-240\Big(167+114\bar l_1+156\bar l_2-81\bar l_3\Big)L_\mu-22680L_\mu^2
\bigg],\notag
\end{align}
where
$\sigma=\sqrt{2}\arctan\frac{1}{\sqrt{2}}$, and for the Adler zero, 
with $\tau=\frac{1}{\sqrt{7}}\arctan\sqrt{7}$,
\begin{align}
s_A^{(2)}
&=\frac{M_\pi^2}{2},
\\
s_A^{(4)}
&=-\frac{M_\pi^2\xi}{144}
\Big[1163+2\Big(107\bar l_1+158\bar l_2-90\bar l_3\Big)
\notag\\
&
-908\tau-4224\tau^2\Big],\notag
\\
s_A^{(6)}
&=-\frac{M_\pi^2\xi^2}{81}
\bigg[
k_A
+\frac{1}{16}\Big(3103\bar l_1^2+7364\bar l_1\bar l_2
+4108\bar l_2^2\notag\\
&-2610\bar l_1\bar l_3-2340\bar l_2\bar l_3
-1620\bar l_3^2\Big)\notag\\
&+\frac{1}{896}\Big(1454303-1390328\tau-4098432\tau^2\Big)\bar l_1\notag\\
&+\frac{1}{896}\Big(2410997-2315912\tau-6820608\tau^2\Big)\bar l_2\notag\\
&-\frac{45}{28}\Big(470-557\tau-984\tau^2\Big)\bar l_3\notag\\
&+216\pi^4\Big(240r_1^r+240r_2^r+428r_3^r
+836r_4^r+1047r_5^r+483r_6^r\Big)\notag\\
&-\frac{1}{128}\Big(21653-204\bar l_1+63084\bar l_2
-12960\bar l_3\Big)L_\mu
-195L_\mu^2
\bigg],\notag
\end{align}
where the constant $k_A=123.562$ contains, besides $\pi^2$ and powers of $\tau$, the transcendental terms $\zeta(3)$, $\log\frac{7}{2}$, and polylogarithms evaluated at $e^{2i\sqrt{7}\tau}$~\cite{Heuser:2024biq}. Since $s_A^{(2)}$ does not depend on $F$, neither $\bar l_4$ nor $r_F^r$ enter $s_A$ to this order. In all cases, we checked explicitly the scale independence, using the $\beta$ functions from Ref.~\cite{Bijnens:1999hw}.

\section{Details of the log-aware inference}
\label{app:bayes}

The prior for the coefficients resulting from the naturalness model of Sec.~\ref{sec:logaware} is
\begin{equation}
\pi(\pmb{\aleph}|\bar{c}^2,\mathcal{I}) =
\frac{1}{(2\pi \bar{c}^2)^{N/2}}\exp\left(-
\frac{\pmb{\aleph}^2}{2\bar{c}^2}
\right)~,
\end{equation}
where $N$ is the number of elements in the vector $\pmb{\aleph}$. We neglect correlations between the coefficients $\aleph_{n,m}$.
Integrating over the width then gives \cite{Melendez:2019izc}
\begin{align}
\label{eq:aleph_prior}
    \pi(\pmb{\aleph}|\nu_0, \tau_0^2 , \mathcal{I})
    &=
    \int_0^\infty d\bar{c}^2 \pi(\pmb{\aleph}|\bar{c}^2, \mathcal{I})\pi(\bar{c}^2|\nu_0,\tau_0^2,\mathcal{I})\\
    &=\frac{\Gamma\Big(\frac{N+\nu_0}{2}\Big)}{\Gamma\Big(\frac{\nu_0}{2}\Big)(\pi\nu_0 \tau_0^2)^{N/2}}
    \left( 1 + \frac{1}{\nu_0\tau_0^2}\pmb{\aleph}^2\right)^{-(\nu_0 + N)/2}.\notag
\end{align}

At a given order, we calculate the Jacobian for the change of variables from ${\bf d}=(d_{1,0},d_{1,1},\ldots)^T$ to $\pmb{\aleph}$,
\begin{equation}
    J = \left| \det \frac{\partial \pmb{\aleph}}{\partial {\bf d}}\right|,
\end{equation}
and express the likelihood for the coefficients ${\bf d}$
\begin{align}
    p({\bf d}|\Lb, \nu_0,\tau_0^2,\mathcal{I})&= J(\Lb)\pi(\pmb{\aleph}|\nu_0,\tau_0^2,\mathcal{I})\\
    &= \frac{J(\Lb)\Gamma\Big(\frac{N+\nu_0}{2}\Big)}{\Gamma\Big(\frac{\nu_0}{2}\Big)(\pi\nu_0 \tau_0^2)^{N/2}}
    \left( 1 + \frac{1}{\nu_0\tau_0^2}\pmb{\aleph}^2\right)^{-(\nu_0 + N)/2}.\notag
\end{align}
The posterior for the naturalness scale $\Lb$ is given by
\begin{equation}
    p(\Lb|{\bf d},\nu_0,\tau_0^2,\mathcal{I})
    = \frac{p({\bf d}|\Lb,\nu_0,\tau_0^2,\mathcal{I})\pi(\Lb)}{\int d\lambda\, p({\bf d}|\lambda,\nu_0,\tau_0^2,\mathcal{I})\pi(\lambda)},
\end{equation}
where we use a log-uniform prior for $\Lb$, i.e.,
\begin{equation}
    \pi(\Lb) \propto \frac{1}{\Lb}.
\end{equation}
Using $N=2$ at NLO, we obtain for the posterior of the naturalness scale
\begin{equation}
    \label{eq:posterior_nlo}
    p(\Lb|{\bf d}_\text{NLO},\nu_0,\tau_0^2,\mathcal{I})
\propto \frac{\rho^2}{\Lb}\left(1+ \frac{\pmb{\aleph}^2}{2}\right)^{-2},
\end{equation}
and using $N=5$ for $F_\pi$ at NNLO,
\begin{equation}
\label{eq:posterior_n2lo}
p(\Lb|{\bf d}_\text{NNLO},\nu_0,\tau_0^2,\mathcal{I})
\propto \frac{\rho^8}{\Lb}\left( 1+\frac{\pmb{\aleph}^2}{2}\right)^{-7/2}.
\end{equation}
The omitted normalization factors in Eqs.~\eqref{eq:posterior_nlo} and \eqref{eq:posterior_n2lo} are evaluated numerically. In both expressions the powers of $\rho$ arise from the Jacobian, while the factor $1/\Lb$ comes from the log-uniform prior for $\Lb$.

\bibliographystyle{apsrev4-1_mod}
\balance
\biboptions{sort&compress}
\bibliography{references}

\end{document}